\documentclass[twocolumn,twocolappendix]{aastex631}
\usepackage[utf8]{inputenc}

\newcommand{\avg}[1]{\langle #1 \rangle}

\usepackage{amsthm}
\usepackage{amssymb}
\usepackage{lineno}
\usepackage{ulem}
 
\usepackage{forest}

\usepackage{savesym}
\savesymbol{tablenum}
\usepackage{siunitx}
\restoresymbol{SIX}{tablenum}

\begin{document}

\title{IOP4, the Interactive Optical Photo-Polarimetric Python Pipeline}

\correspondingauthor{Juan Escudero Pedrosa}
\email{jescudero@iaa.es}

\author[0000-0002-4131-655X]{Juan Escudero Pedrosa}
\affiliation{Instituto de Astrofísica de Andalucía, Glorieta de la Astronomía, s/n. Granada, 18008, Spain}

\author[0000-0002-3777-6182]{Iván Agudo}
\affiliation{Instituto de Astrofísica de Andalucía, Glorieta de la Astronomía, s/n. Granada, 18008, Spain}

\author[0000-0001-9400-0922]{Daniel Morcuende}
\affiliation{Instituto de Astrofísica de Andalucía, Glorieta de la Astronomía, s/n. Granada, 18008, Spain}

\author[0000-0002-4241-5875]{Jorge Otero-Santos}
\affiliation{Instituto de Astrofísica de Andalucía, Glorieta de la Astronomía, s/n. Granada, 18008, Spain}

\author[0000-0003-2464-9077]{Giacomo Bonnoli}
\affiliation{INAF Osservatorio Astronomico di Brera, Via E. Bianchi 46. Merate (LC), 23807, Italy}

\author{Vilppu Piirola}
\affiliation{Department of Physics and Astronomy, University of Turku, Vesilinnantie 5. Turku, FI-20014, Finland}

\author[0000-0001-8286-5443]{César Husillos}
\affiliation{Geological and Mining Institute of Spain (IGME-CSIC), Calle Ríos Rosas 23, E-28003, Madrid, Spain}
\affiliation{Instituto de Astrofísica de Andalucía, Glorieta de la Astronomía, s/n. Granada, 18008, Spain}

\author{Mabel Bernardos}
\affiliation{Instituto de Astrofísica de Andalucía, Glorieta de la Astronomía, s/n. Granada, 18008, Spain}

\author[0000-0002-3882-9477]{Rubén López-Coto}
\affiliation{Instituto de Astrofísica de Andalucía, Glorieta de la Astronomía, s/n. Granada, 18008, Spain}

\author{Alfredo Sota}
\affiliation{Instituto de Astrofísica de Andalucía, Glorieta de la Astronomía, s/n. Granada, 18008, Spain}

\author{Victor Casanova}
\affiliation{Instituto de Astrofísica de Andalucía, Glorieta de la Astronomía, s/n. Granada, 18008, Spain}

\author{Fran J. Aceituno}
\affiliation{Instituto de Astrofísica de Andalucía, Glorieta de la Astronomía, s/n. Granada, 18008, Spain}

\author{Pablo Santos-Sanz}
\affiliation{Instituto de Astrofísica de Andalucía, Glorieta de la Astronomía, s/n. Granada, 18008, Spain}

\begin{abstract}

IOP4 is a pipeline to perform photometry and polarimetry analysis of optical data from Calar Alto (CAHA) and Sierra Nevada (OSN) observatories. IOP4 implements Object Relational Mapping (ORM) to seamlessly integrate all information about the reduction and results in a database which can be used to query and plot results, flag data and inspect the reduction process in an integrated fashion with the whole pipeline. It also ships with an already built-in web interface which can be used out of the box to browse the database and supervise all pipeline processes. It is built to ease debugging and inspection of data. Reduction from five different instruments are already implemented: RoperT90, AndorT90 and DIPOL (at OSN 0.9m telescope), AndorT150 (OSN 1.5m telescope) and CAFOS (CAHA 2.2m telescope). IOP4's modular design allows for easy integration of new observatories and instruments, and its results have already featured in several high-impact refereed publications. In this paper we describe the implementation and characteristics of IOP4.

\end{abstract}

\keywords{Data analysis --- Photometry --- Polarimetry -- Astronomy databases -- Open source software}

\section{Introduction} \label{sec:intro}

Optical photo-polarimetric observational programs, especially those dedicated to monitoring, can regularly produce large quantities of data that can take considerable time and effort to be managed and reduced. The effort necessary to produce good quality results extends beyond the use of automatic tools and can include a human-supervised iterative process of debugging the reduction and comparing the employed methods and results with those of different programs. The use of automatic tools is therefore necessary, however, in many instances they obscure the process of reduction and intermediate results, making the debugging of any problem in the results a hard task.

IOP4 implements Object Relational Mapping using Django's ORM system. This allows to transparently keep the database schema up to date with the models used by the pipeline without any need to mess with the underlying SQL queries. The choice of Django's as ORM backend allows to seamlessly use the rest of Django Framework to serve the results, including but not limited to its admin interface to inspect the database, and Django's debug web server. These tools are all written in Python and packaged for pip and conda, a programming language and distributions that many astronomers are already familiarized with. This also makes IOP4 a multi-platform software, compatible both with macOS and Linux.

IOP4 builds on top of existing technologies, some of them already cited: Django (ORM and web application framework), SQLite (default database backend), astrometry.net (blind astrometric calibration, \citealt{Lang:2010}), Bokeh (high quality and interactive plots in the web interface), Vue.js (single page web application), Quasar (user interface components) and JS9 (a ds9 web port for interactive FITS visualization). Many other open-source packages are used, the complete list can be found in the \texttt{setup.toml} file. All of them are automatically installed together with IOP4.

Hardware requirements of IOP4 are low, even compared today to modern day consumer-end laptops. The blind astrometric calibration using the astrometry.net solver and its index files can take as much as 35GB with the default configuration, although this requirement can be lowered. SSD storage is also recommended, since it needs to read significant amounts of data especially for the astrometric calibration. Although it is able to run on a single-core, significant speed improvements can be gained from using up to 20 cores, the recommended maximum with the default configuration and database (DB) backend. Higher levels of concurrency are possible by tuning the DB configuration. A good internet connection can speed up the first execution of IOP4, including tests, since it will need to download astrometry index files.

IOP4 facilitates both automatic reduction and manual inspection of the procedure. The \texttt{iop4} command provides several options to select different epochs and files and automatically process them. The \textsc{iop4lib} allows any user to write its own custom reduction scripts, and provides the tools to invoke any part of the reduction procedure and inspect and manipulate any object in the database from an interactive terminal, a Python script or a Jupyter Notebook. The \textsc{iop4api} implements several API (Application Public Interface) endpoints and web applications that allow end users to easily query results, produce plots, flag bad data and inspect any object. The portal can be used standalone as in the ready for use \textsc{iop4site} project, or integrated into other sites to be served to the general public.

\section{General reduction procedure}

The automatic reduction procedure generally starts with a simple invocation of the \texttt{iop4} command. The main script can be requested to select epochs in the local archive or to discover and download epochs in the remote telescope repositories. For each epoch, the script goes through the usual steps of photo-polarimetric reduction: 

\begin{enumerate}
    \item Classification of raw science images: This includes their type (bias, darks, flats and science images), discovering the instrument and type of observation (photometry or polarimetry), and translation of standard and non-standard keywords (exposure time, band, datetime, rotator angles, objects).
    \item Creation of master calibration frames: Images of each type (bias, darks and flats) are grouped and merged together to create the master calibration nights available for each night.
    \item Reduction and calibration of science images: Raw images are applied the corresponding master calibration frames. The reduced images are also given a correct World Coordinate System (WCS) in their header after astrometric calibration.
    \item Computation of photo-polarimetric results: At the moment, the procedure implements relative photometry using known calibrators in the field, and polarimetry both for half-wave ($\lambda/2$) retarder based polarizers and polarized filter wheels.
    \item Post-processing of results: This includes correction of magnitude and degree of polarization to account for the host contribution (\citealt{Nilsson:2007}), and the possibly needed transformation to a standard photometric system, if the instrument used for the observations does not have one.
\end{enumerate}

\section{Data organization}

\begin{figure}[htbp!]
    \includegraphics[width=\linewidth, height=\textheight, keepaspectratio]{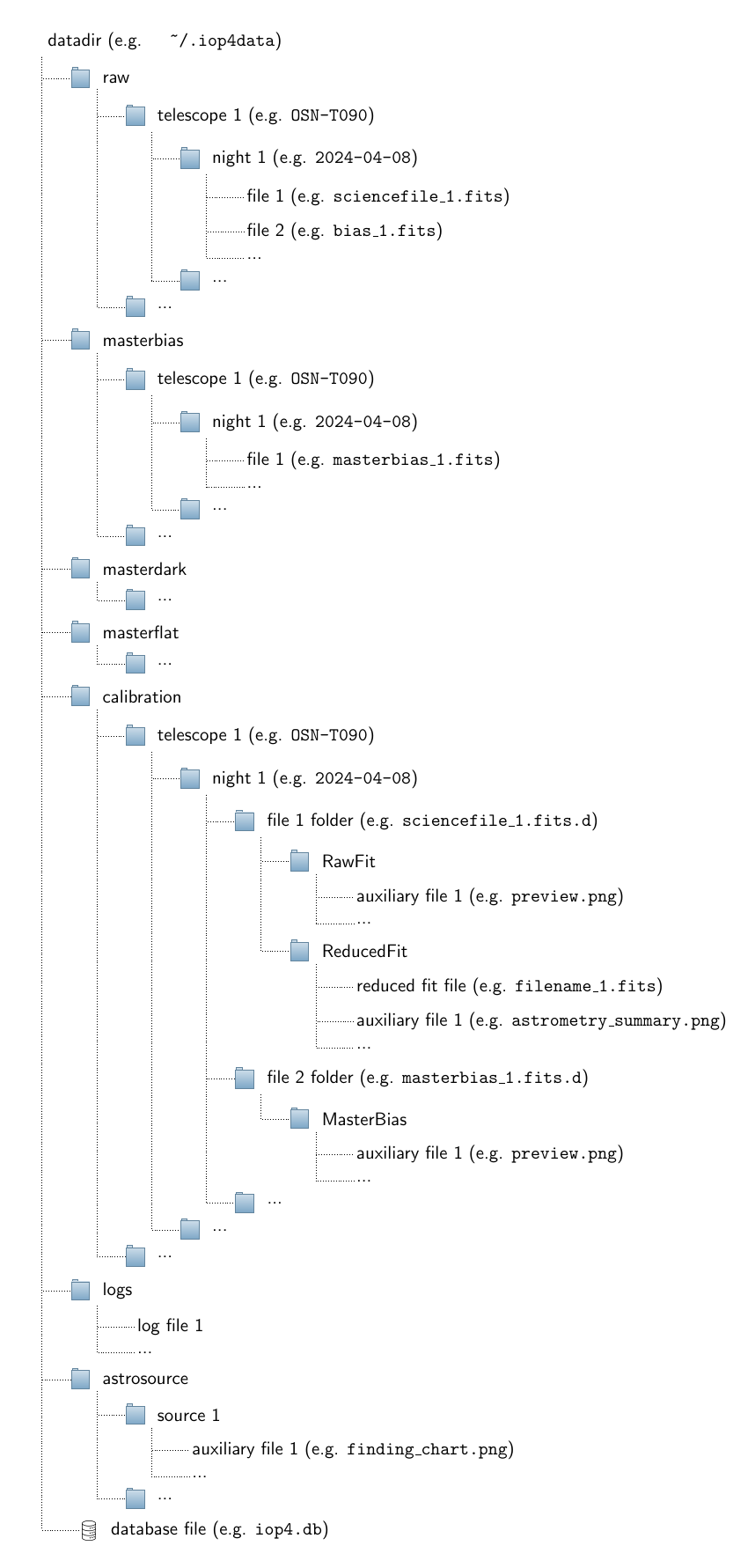}
    \caption{Directory structure of the IOP4 local archive.}
    \label{fig:dir_struct}
\end{figure}

IOP4 data directory structure follows the typical hierarchical schema shown in Fig. \ref{fig:dir_struct}. In this schema, all raw data is stored and isolated under a single folder (\texttt{raw/}), which allows to set up a local archive of the original data without any modifications for long-term conservation, and to set up the necessary permissions to protect and share it with other users independently of the rest of IOP4-created files. Under the raw directory, data is organized first by telescope and then by night of observation. Other files such as built master calibration frames and reduced images are stored separately. Also auxiliary images such as automatically built previews, finding charts, summary plots, etc, which are too heavy to be stored in a database, are stored under different folders.

\begin{figure*}[htbp!]
    \centering
    \includegraphics[width=1.0\linewidth]{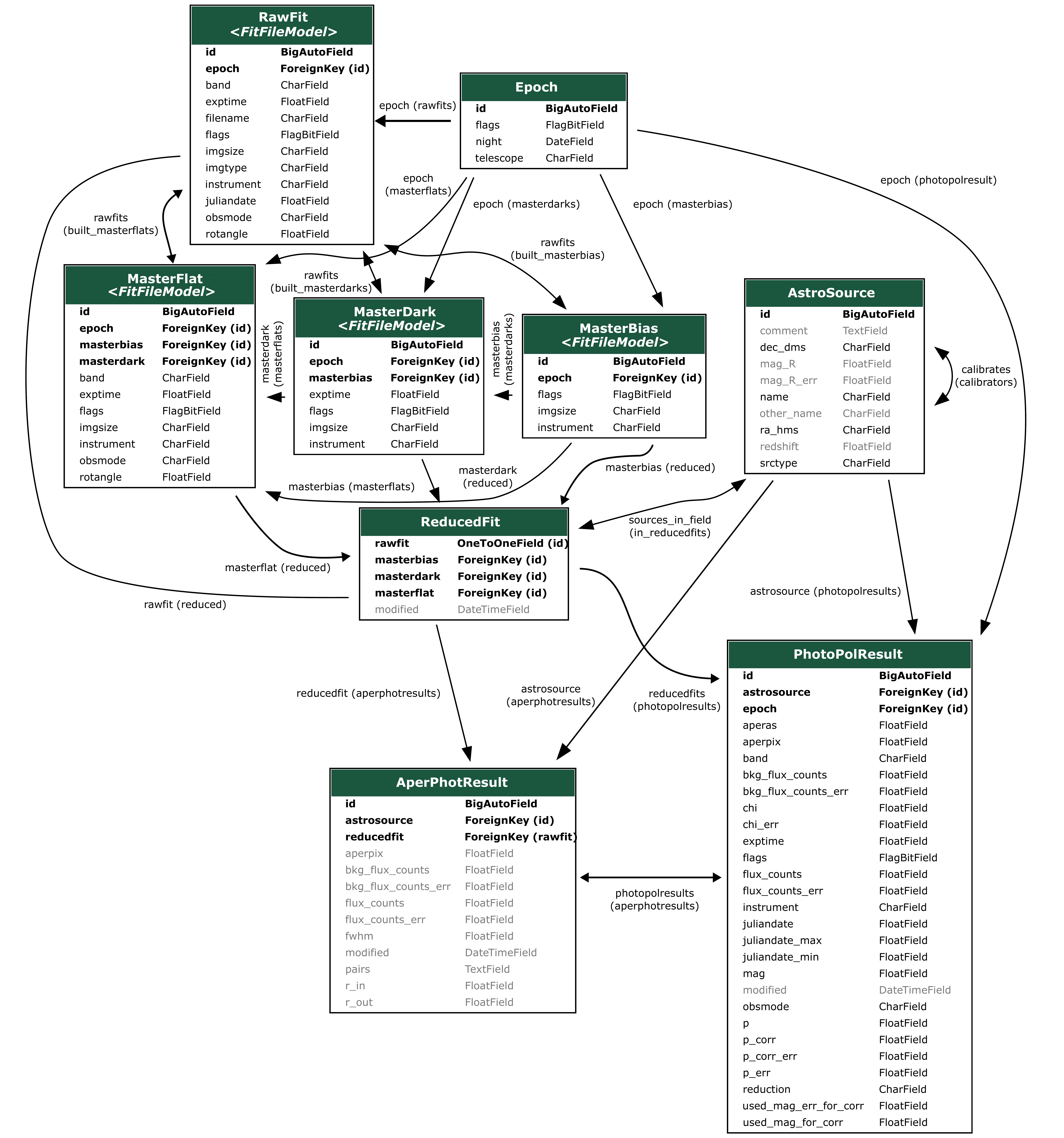}
    \caption{Database schema of IOP4. \textit{ForeignKey} relationships relate one or several instances of model A to a single instance of model B. In \textit{ManyToMany} relationships, multiple instances of model A and B are linked together. The latter does not show a field in the tables of this diagram, since the relationship is established through a hidden table omitted here. Some fields of the \texttt{AstroSource} model have been also omitted to save space (corresponding to literature magnitudes in other bands, e.g. \texttt{mag\_B}, etc). An arrow in any direction signify multiple instances being linked to the origin (e.g. several RawFit(s) are linked to one Epoch).}
    \label{fig:iop4api_models_simple}
\end{figure*}

The database schema is shown in Fig. \ref{fig:iop4api_models_simple}. IOP4 implements object-relational mapping (ORM). In ORM, objects in a object-oriented programming language (Python classes) are mapped to tables in a relational database, while instances of these objects correspond to rows in each of the tables. This enormously simplifies the interaction with the database, removing the need for writing SQL queries and manipulating the database to keep its structure and content updated.

The default database backend is {SQLite}\footnote{\url{https://www.sqlite.org/}.}, a in-process library that implements a self-contained, serverless, zero-configuration, and transactional SQL database; and uses the Django ORM\footnote{\url{https://docs.djangoproject.com/}.} to interact with it. 
Customizations and changes to the DB schema can be written in the code (e.g, adding an attribute to the photo-polarimetric result model) and they will be automatically propagated to the DB schema by the Django migration system.

\section{Astrometric calibration}

Astronomical images are usually distributed in the FITS format (\citealt{FITS:III}). The format allows for a World Coordinate System (WCS, \citealt{WCS:I}) to be incorporated in the metadata or header section of the FITS file. However, most raw science images from telescopes do not include this precise information and need to be calibrated. For most images, which have a wide field of view ($>\SI{7}{\arcmin}$) this is done by a local solver%
\footnote{%
    The local solver is integrated as an external Python module dependency in the \texttt{pyproject.toml} file and is installed next to IOP4 automatically. The module is a wrapper around the native C functions of the astrometry.net library.
    \url{https://github.com/neuromorphicsystems/astrometry}}
 based on the astrometry.net library solver\footnote{\url{https://github.com/dstndstn/astrometry.net}}. 

\begin{figure}[htbp!]
    \centering
    \includegraphics[width=0.9\linewidth]{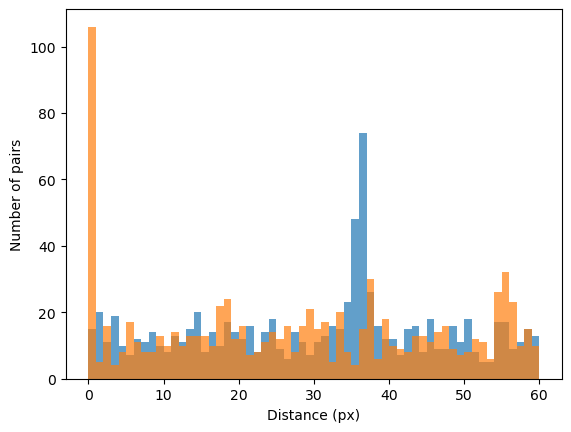}
    \caption{Distribution of the distances between all possible pairs in a imaging polarimetric frame. The two peaks correspond to the distance between the ordinary and extraordinary sources in the image. The image and the corresponding found pairs can be found in Fig. \ref{fig:pairs_paired_example}. The orange and blue distributions correspond to the distances in X and Y axis respectively. The image corresponds to a CAFOS photo-polarimetric observation of the BL Lacertae field.}
    \label{fig:pairs_distance_distribution}
\end{figure}

\begin{figure}[htbp!]
    \centering
    \includegraphics[width=0.9\linewidth]{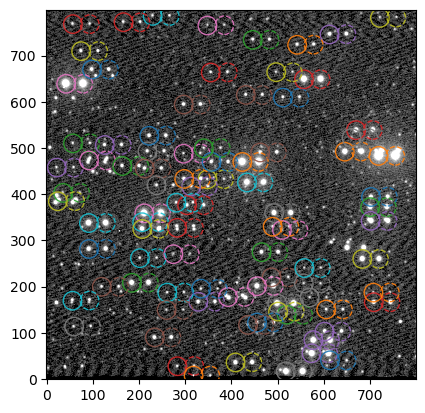}
    \caption{Paired sources in an example of CAFOS imaging polarimetric observation of the BL Lacertae field. The distance used for pairing was obtained from Fig. \ref{fig:pairs_distance_distribution}. The resulting calibrated image can be found in Fig. \ref{fig:pairs_calibrated_example}. Ordinary and Extraordinary sources are circled with the same color.}
    \label{fig:pairs_paired_example}
\end{figure}

For observations with Ordinary (O) and Extraordinary (E) images (e.g. CAFOS and DIPOL imaging polarimetric observations), astrometric calibration involves a previous step of separating the detected sources in pairs. The source pairing is done by finding the most common distance between all pairs of sources in the image, which results in two distributions like those in Fig. \ref{fig:pairs_distance_distribution}. Then, the separation can proceed by looking at which pairs are at the right distance (Fig. \ref{fig:pairs_paired_example}). 
Without any constraint, this process is not completely error free, specially for images with few detected sources. However, since the distance between pairs is usually a stable property of the instrument, the known distance between pairs (obtained by the unconstrained pairing) for an instrument can be used as initial input, which improves the success rate up to \SI{100}{\percent}.

\begin{figure}[htbp!]
    \centering
    \includegraphics[width=0.8\linewidth]{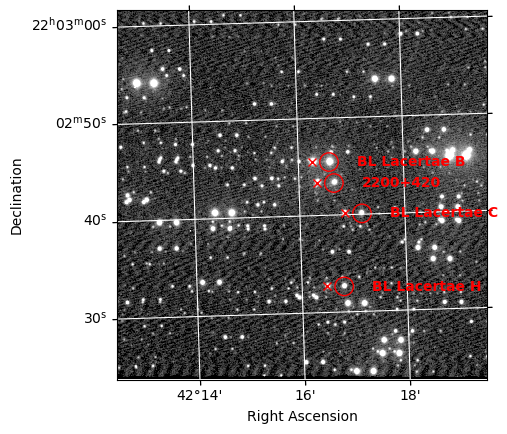}
    \caption{Calibrated CAFOS photo-polarimetric image of the BL Lacertae field. The positions of the ordinary image of BL Lacertae (labeled as 2200+420) and its calibrators (B, C, H) are indicated with a circle. The position of the extraordinary images are represented with an 'x'.}
    \label{fig:pairs_calibrated_example}
\end{figure}

The detected sources in the image, or the ordinary set of pairs for images with pairs, can then be used as input for the local astrometry solver (\citealt{Lang:2010}).
The solver compares invariant hashes computed from the detected source positions to pre-computed hashes from astronomical catalogs. Although it does not require an initial guess of the position nor the pixel size (blind solving), the speed and accuracy of the solution is greatly improved by providing the known pixel size for the instrument and a hint position, obtained from the header of the images.
It returns a list of matches and their corresponding log-odds.
Its success is dependent on the source detection step, and therefore several attempts are made with different detection parameters (such as signal threshold) until a good match is found.
The resulting WCS is written to the header of the reduced FITS file. The WCS for the second or extraordinary set of pairs is directly built from the first one by translation, and written next to it in the same header. An example of the result can be found in Fig. \ref{fig:pairs_calibrated_example}, which corresponds to the pairs in Figs. \ref{fig:pairs_distance_distribution} and \ref{fig:pairs_paired_example}.

\subsection{A quad hash for DIPOL polarimetry images}
\begin{figure*}[htbp!]
    \centering
    \includegraphics[width=0.45\linewidth]{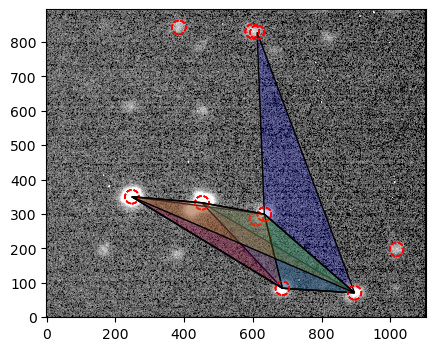}
    \includegraphics[width=0.45\linewidth]{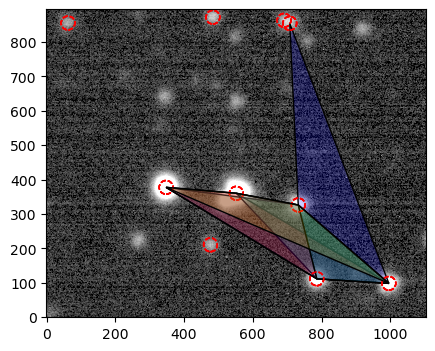}
    \caption{%
    DIPOL polarimetry (left) and photometry (right) frames of BL Lacertae observations. 
    Only the shown subframe ($\SI{2.5}{\arcminute} \times \SI{2.0}{\arcmin}$) is usually saved to disk during polarimetric observations with DIPOL, which prevents blind solving of the image. The full photometry field ($\SI{9.2}{\arcmin}\times\SI{6.3}{\arcmin}$) is saved, however, and the correspondence between both images can be found by comparing quads of sources through a hashing algorithm. This allows us to automatically distinguish the O and E images of our target source in the field.}
    \label{fig:dipol_quads_comp}
\end{figure*}

To reduce the disk capacity requirements of DIPOL polarimetry images, only a subframe of the full field of DIPOL camera is saved (Sect. \ref{sec:inst:dipol}). The reduced field of view does not usually contain enough stars to perform the astrometric calibration using the default solver. In fact, in many cases, only the O and E images of the target source are visible in the image. For the case of images in which only 1-2 pairs of sources appear, the astrometric calibration can be as simple as using the central pair as a reference for the creation of the WCS with the known angle of the RA-DEC grid. However, this process is error-prone when more than two pairs of sources appear in the image, as the likelihood of choosing the wrong pair increases. Even for the case of high number of pairs, the size of the subframe ($\SI{2.5}{\arcminute} \times \SI{2.0}{\arcmin}$) is too small even for the smallest skymarks of the default astrometry.net index files\footnote{\url{http://astrometry.net/doc/readme.html}}. The problem of choosing the right pair of sources as the reference can be solved by comparing the subframe of the polarimetry observation with the central part of a calibrated photometry (full frame) field. To this end, we have implemented a hash algorithm loosely based on \cite{Lang:2010}. The proposed hash is invariant under rotation and reflection, although not under scaling. An example of this algorithm at work can be found in Fig. \ref{fig:dipol_quads_comp}. It takes the ten brightest sources in each image (including E and O images) and compares all the quads between them. The best matching quads are used for identifying the target star.

\section{High-level web interface}

As one of its main goals, IOP4 also provides an API to interact with the data and a web interface to act as a client. Both are provided by the \textsc{iop4api} Django application. It defines API end points to query results, produce interactive web-based plots and flag data. It also integrates the \textsc{iop4admin} site, a customized Django-admin. The admin site allows us to inspect any models in the database (Fig. \ref{fig:iop4api_models_simple}).

The IOP4 catalog is part of the database. IOP4 uses a single, unified catalog that gathers all information about the sources of interest, including the target sources, calibrators, relationships between them, comments for observers, etc. Editing of the catalog can be done through IOP4 as with any other model, or through the web interface, and the changes take immediate effect on the reduction process of the pipeline.

\begin{figure*}
    \centering
    \includegraphics*[width=0.9\textwidth]{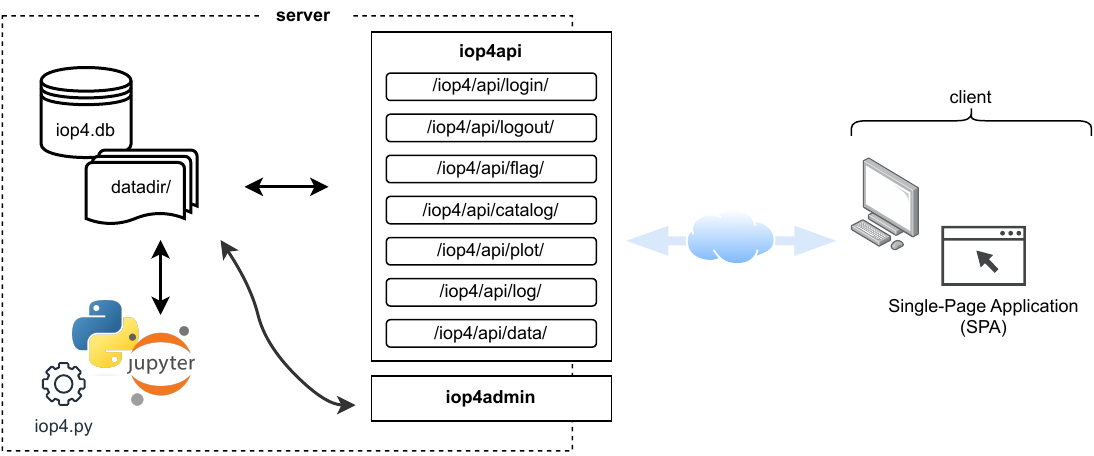}
    \caption{Different ways of interacting with IOP4. The pipeline (through the \textsc{iop4} command, Python script or Jupyter notebooks) can directly interact with the data aided by \textsc{iop4lib}. The \textsc{iop4api} application exposes a series of API endpoints that allow the client to authenticate itself, query, plot and flag data, and explore catalog and logs from a web browser through a Single-Page Application (SPA). The \textsc{iop4admin} site independently provides a way to interact and edit all models in the database.}
    \label{fig:interactions}
\end{figure*}

The situation is illustrated in Fig. \ref{fig:interactions}. The exposed API endpoints are used by the single-page web application (SPA) to authenticate the client, query, plot and flag data, and explore logs and catalog. The SPA provides a viewer for the colored logs and allows filtering by logging level (debug, info, warning, and error) and string searching. The interactive plot is built and serialized in the server using the  Bokeh\footnote{\url{https://docs.bokeh.org/en/latest/}} (\citealt{Bokeh}) Python library, sent to the client and rendered by the BokehJS library. The interactive plot can be used to directly flag the data. TabulatorJS\footnote{\url{https://tabulator.info/}} is used to display the results, and allows filtering, selecting columns and exporting to several data formats such as CSV. The SPA itself is built in html, css and javascript using Vue.js framework as a standalone script to avoid the build step and allow IOP4 to be installed and used by the astrophysics community in a familiar way (through pip or conda).

The web application provides an auxiliary tool to facilitate the addition to the catalog of calibrators for sources with no known previously documented calibrators.
A search for standard stars with constant brightness is performed within the PanSTARRS\footnote{\url{https://catalogs.mast.stsci.edu/panstarrs/}} catalog. For this, we filter stars within the FoV of all instruments that have a relatively large amount of observations available (typically $N \geq 10$), with a standard deviation of their aperture SDSS $\mathrm{gri}$ magnitudes $< 0.01$. In order to use as calibrators stars that do not have the risk of saturating the images, we also restrict the search to targets with magnitudes between \num{13} and \num{18}, typically. With these considerations, we retrieve the $\mathrm{gri}$ aperture magnitudes of non-variable stars that will be used as calibrators.
Due to the different photometric system filter used by the PanSTARRS database (based on $grizy$ filters) and that from the instruments implemented in IOP4 (generally equipped with standard Johnson-Cousins filters) a conversion between photometric systems is needed. Three transformations are currently implemented in IOP4: \cite{Jordi:2006}, \cite{Jester:2005} and \cite{Lupton:2005}. As explained by these authors, these transformations are suitable for stars, with the caveat of \cite{Jester:2005} being only suitable for stars with $R_{C}-I_{C}<1.15$. All three transformations have been found to be compatible for the calibrators added following this procedure. IOP4 implements by default the transformations from \cite{Lupton:2005}.

\section{Current instruments}

Five instruments from three different telescopes are already implemented in IOP4: RoperT90, AndorT90 and DIPOL (at OSN 0.9m telescope), AndorT150 (OSN 1.5m telescope) and CAFOS (CAHA 2.2m telescope). The modular design of IOP4 allows easily integrating new instruments and observatories. The \texttt{Telescope} base class provides the skeleton over which to implement new telescopes. It implements the necessary methods to query, download (e.g. through the \texttt{FTPArchiveMixin} class) and perform the initial classification of observing epochs.

Integrating a new instrument is as easy as subclassing the \texttt{Instrument} subclass. The new subclass must provide the necessary information to identify the instrument, and possibly implement methods to translate non-standard keywords (if any), extract position and size hints for astrometry or perform a custom reduction process.

\subsection{RoperT90, AndorT90, AndorT150} \label{sec:inst:osn_ccds}
IOP4 implements data reduction from the current CCD (Andor ikon-L) cameras at the 0.9m and 1.5m telescopes in Sierra Nevada Observatory (OSN). It also implements the old Roper (VersArray) cameras (which were installed until October 2021 and July 2018 respectively). The AndorT90 instrument mounted at the Nasmyth east focus of the T090 telescope has a $\SI{13.2}{\arcminute} \times \SI{13.2}{\arcminute}$ field of view and pixel size of $\SI{0.387}{\arcsecond {px}^{-1}}$. The AndorT150 instrument at the Nasmyth west focus of the T150 has similar characteristics, with a $\SI{7.92}{\arcminute} \times \SI{7.92}{\arcminute}$ field of view and pixel size of $\SI{0.232}{\arcsecond {px}^{-1}}$. They are cooled down down to \SI{-80}{\celsius} without need of liquid nitrogen, and can be further cooled down to \SI{-100}{\celsius} using liquid refrigerant. The low temperatures make the use of dark current calibration frames unnecessary. Apart from the usual band filter wheels, a polarized filter wheel is available at the T090 and the T150 that allows polarimetry measurements by taking series of four images at varying polarized angles in \SI{45}{\deg} steps. From these, the total flux and raw Stokes parameters are computed as
\begin{equation}
        F = \avg{f} = \frac{1}{4}\sum_i{f_i}
\end{equation}
and
\begin{eqnarray}
    q_\text{raw} &=& \frac{ f_{0} - f_{90} }{ f_{0} + f_{90} }	\\
    u_\text{raw} &=& \frac{ f_{45} - f_{-45} } { f_{45} + f_{-45} }
    ~,
\end{eqnarray}
where $f_i$ are the fluxes at each rotator angle. The instrumental polarization is corrected by applying an offset 
\begin{eqnarray}
    q_c &=& q_\text{raw} - q_{\text{inst}}	\\
    u_c &=& u_\text{raw} - u_{\text{inst}}
\end{eqnarray}
and a rotation
\begin{eqnarray}
    q &=& q_c \cos{ \left( 2\chi_{\text{inst}} \right) } - u_c \sin{ \left( 2\chi_{\text{inst}} \right) } 	\\
    u &=& u_c \sin{ \left( 2\chi_{\text{inst}} \right) }  + u_c \cos{ \left( 2\chi_{\text{inst}} \right) }		
    ~,
\end{eqnarray}
from which the linear polarization degree and polarization angle are computed as
\begin{eqnarray}
    p &=& \sqrt{q^2 + u^2}	 
    \label{eq:p_qu}	\\
    \chi &=& \frac{1}{2} \arctan{\left(u,q\right)}  	
    \label{eq:chi_qu}
    ~.
\end{eqnarray}

\subsection{CAFOS} \label{sec:inst:cafos}
The CAFOS (Calar Alto Faint Object Spectrograph) instrument, mounted on the 2.2m telescope at Calar Alto Observatory, provides imaging polarimetry capabilities. It is equipped with a Wollaston prism and a rotable $\lambda/2$ retarder plate that provides two polarized images separated \SI{18}{\arcsecond}. 
Only a $\num{800}\times\num{800}$ subframe of the full $\num{2048}\times\num{2048}$ CCD chip is typically used, with a field of view of $\SI{34}{\arcminute}\times\SI{34}{\arcminute}$ and a pixel size of $\SI{0.530}{\arcsecond {px}^{-1}}$. The CCD chip is cooled to temperatures lower than \SI{-100}{\celsius}, making the dark current negligible. Several filters are available, while polarimetry observations are usually taken in Johnson R. The ordinary (O) and extraordinary (E) images at each angle are used to compute the total flux and the reduced Stokes parameters as (\citealt{zapatero_caballero_bejar:2005})
\begin{eqnarray}
    F = \frac{1}{N} \sum_i \frac{f_{E,i} + f_{O,i}}{2}
\end{eqnarray}
and
\begin{eqnarray}
    R_Q &=& \sqrt{\frac{ f_{O,0} / f_{E,0} }{ f_{O,45} / f_{E,45} }} 	\\
    R_U &=& \sqrt{\frac{ f_{O,22} / f_{E,22} }{ f_{O,67} / f_{E,67} }} 
    ~,
\end{eqnarray}
where $f_{O,i}$ and $f_{E,i}$ are the fluxes of the ordinary and extraordinary images of the source at each angle $i$. From these, the Stokes parameters are reconstructed as
\begin{eqnarray}
    Q_I &=& \frac{R_Q-1}{R_Q+1}	\\
    U_I &=& \frac{R_U-1}{R_U+1}
    ~,
\end{eqnarray}
and the polarization degree and polarization angles from Eqs. \ref{eq:p_qu} and \ref{eq:chi_qu}. While instrumental polarization is negligible for CAFOS, the polarization angle still needs to be corrected according to
\begin{equation}
    \chi = \chi - CPA 
    \label{eq:chi_CPA}
    ~,
\end{equation}
where CPA is the Zero Polarization Angle.

\subsection{DIPOL} \label{sec:inst:dipol}

The DIPOL-1 polarimeter is thoroughly described in \cite{dipol:Piirola:2020} and \cite{dipol:Jorge:2024}. It is based on a $\lambda/2$ retarder plate attached to a rotator and a high readout speed CMOS camera. Installed at the OSN-T090 the instrument has a field of view of $\SI{9.2}{\arcmin}\times\SI{6.3}{\arcmin}$ and a pixel scale of $\SI{0.134}{\arcsecond {px}^{-1}}$. Cycles of 16 images are typically taken with varying rotator angles at $\SI{22.5}{\degree}$ steps. For polarimetry observations, usually only a subframe of size $\SI{2.5}{\arcminute} \times \SI{2.0}{\arcmin}$ is saved, greatly reducing used disk space. Dark current calibration frames are necessary. High throughput sharp cutoff R, G and B filters are available. The computation of the polarimetric results is done following \cite{patat:2006}. The Stokes parameters are computed as
\begin{eqnarray}
    Q_\text{raw} &=& \frac{2}{N} \sum_i F_i \cos{\left( \frac{\pi}{2} i \right)}	
    \\
    U_\text{raw} &=& \frac{2}{N} \sum_i F_i \sin{\left( \frac{\pi}{2} i \right)} 	
    ~,
\end{eqnarray}
where the coefficients $F_i$ are computed as
\begin{equation}
    F_i = (F_{O,i} - F_{E,i}) / (F_{O,i} + F_{E,i})	
    ~,
\end{equation}
$F_{O,i}$ and $F_{E,i}$ being the fluxes of the ordinary and extraordinary images of the source for each rotator angle $i$. The instrumental polarization is corrected with an offset
\begin{eqnarray}
    Q &=& Q_\text{raw} - Q_\text{inst}	
    \\
    U &=& U_\text{raw} - U_\text{inst}	
    ~.
\end{eqnarray}
Then, $p$ and $\chi$ are obtained per Eqs. \ref{eq:p_qu}, \ref{eq:chi_qu} and \ref{eq:chi_CPA}. Photometric (full frame) observations are also performed. The conversion of the magnitudes in the Baader $\mathrm{RGB}$ filters to standard Johnson-Cousins $\mathrm{UBVR_cI_c}$ will be treated in a separate paper (in preparation).

\section{Development and CI}

IOP4 is open-source, its code hosted at \url{https://github.com/juanep97/iop4}. 
The repository contains all the necessary code to run IOP4 both as a program (\texttt{iop4} script) and as a library (\textsc{iop4lib}), plus the \textsc{iop4api} Django application and the customized \textsc{iop4admin} site. Both can be readily used through the Django debug server and the provided \textsc{iop4site}, or they can be integrated into another site and deployed for public use (see \textit{Serving IOP4 in production} in IOP4 documentation).
To ease the development process and the maintainability of the code, we implemented Continuous Integration and Continuous Deployment (CI/CD) workflows to perform automatic testing, build the documentation and facilitate the delivery of IOP4.
The source code includes a test suit, the test dataset is freely available at \url{https://vhega.iaa.es/iop4/}.
The size of the astrometry.net solver index files makes inviable using GitHub-hosted ordinary runners for CI. Instead, a self-hosted runner provided by the VHEGA\footnote{\url{https://vhega.iaa.es/}} group with self-provisioning capabilities using \textsc{garm}\footnote{\url{https://github.com/cloudbase/garm}} automatically runs tests on pull requests (PR) and merge commits to the main branch on isolated containers that already provide the test dataset and astrometry index files. Anyone can run the test locally in their computers using \textsc{pytest}\footnote{\url{https://docs.pytest.org/}}.
The documentation can be built using \textsc{Sphinx}\footnote{\url{https://www.sphinx-doc.org/}}, and contains several notebook examples. \textsc{MyST-NB}\footnote{\url{https://myst-nb.readthedocs.io/}} and \textsc{Jupytext}\footnote{\url{https://jupytext.readthedocs.io/}} allow using the percent format for the notebooks, an human-readable format easily integrable with version control software. The example notebooks are automatically run when building the documentation and use the test dataset. The documentation is also automatically built and deployed to GitHub Pages as part of the CI. New releases of IOP4 are automatically deployed to the public PyPi software repository\footnote{\url{https://pypi.org/project/iop4/}}.

\section{Conclusions}

IOP4 is an open source, interactive photo-polarimetric pipeline written in Python.
Its results have already featured in a number of refereed and high impact publications (e.g.
\cite{2023ApJ...942L..10M,2023ApJ...948L..25P,2023ApJ...953L..28M,2023ApJ...959...61E,2023NatAs...7.1245D,2023arXiv231011510M,2024AA...681A..12K,dipol:Jorge:2024}; among others), following the footsteps of its predecessor IOP3\footnote{{\cite{iop3}}} (e.g. \cite{2022Natur.611..677L}, etc), and has provided data to many on-going studies. 
Most of this data comes from the MAPCAT (\citealt{mapcat}) and TOP-MAPCAT programs, having reduced more than 500GB of data from the these programs as of February 2024. The former was running at Calar Alto from 2007 to 2018, the latter has been running from 2018 to the present day both at OSN and CAHA. Both programs are focused on the photo-polarimetric monitoring of blazars, combining regular observations with targets of opportunity (ToO). %
Aided by its parallel processing capabilities, it routinely downloads, reduces and serves results from a full night of observation of %
these programs %
in less than half an hour. Its speed makes it suitable to be used as a real-time analysis tool. Moreover, IOP4 has contributed to the publication of several Astronomer Telegrams (such as \citealt{2023ATel16305....1O, 2023ATel16360....1O}) thanks to the promptness of its results and the ease of use.

Development of IOP4 is on-going. Future releases of IOP4 might include several other instruments and reduction methods. As of version \texttt{v1.1.2}, IOP4 provides a prefabricated night summary script that sends the results of observations to subscribed users. Future versions might include a more advanced alert and trigger system. In any case, the \textsc{iop4lib} already allows the users to create their own scripts, for alerts or any other purpose.

The project welcomes any interested user to participate in IOP4 development and request or contribute to the implementation of new instruments and features.

\begin{acknowledgments}
The IAA-CSIC team acknowledges financial support from the Spanish "Ministerio de Ciencia e Innovación" (MCIN/AEI/ 10.13039/501100011033) through the Center of Excellence Severo Ochoa award for the Instituto de Astrofísica de Andalucía-CSIC (CEX2021-001131-S), and through grants PID2019-107847RB-C44 and PID2022-139117NB-C44.
P.S-S. acknowledges financial support from the Spanish I+D+i project PID2022-139555NB-I00 (TNO-JWST) funded by MCIN/AEI/10.13039/501100011033.
Based on observations made at the Sierra Nevada Observatory (OSN), operated by the Instituto de Astrofísica de Andalucía (IAA-CSIC), and at the Centro Astronómico Hispano-Alemán (CAHA), operated jointly by Junta de Andalucía and the IAA-CSIC.
Development of this software would not have been possible without the invaluable and selfless contributions of the open source community.
\end{acknowledgments}

\vspace{5mm}
\facilities{Instituto de Astrofísica de Andalucía (IAA-CSIC), Observatorio de Sierra Nevada (OSN), Observatorio de Calar Alto (CAHA)}

\software{
    Django (\citealt{django}),
    SQLite (\citealt{sqlite}),
    Vue.js (\citealt{vue}),
    numpy (\citealt{numpy}),
    astropy (\citealt{astropy:2013,astropy:2018,astropy:2022}),
    photutils (\citealt{photutils}),
    astrometry.net (\citealt{Lang:2010})
}

\nocite{*}
\bibliography{citations,feat_iop4}{}
\bibliographystyle{aasjournal}

\end{document}